\documentclass[prl,reprint,superscriptaddress,amsmath,amssymb,aps]{revtex4-2}

\usepackage{graphicx}
\usepackage{dcolumn}
\usepackage{bm}
\usepackage[section]{placeins}
\usepackage{bibunits}

\makeatletter
\def\NAT@bibsetnum#1{%
  \setlength{\topsep}{\z@}%
  \NATx@bibsetnum{#1}%
}
\def\endthebibliography{%
  \auto@bib@innerbib
  \endNAT@thebibliography
  \aftergroup\auto@bib@empty
}
\makeatother

\DeclareMathOperator{\sgn}{sgn}

\begin{document}

\title{An Absorbing State Model with an Emergent Ising Transition}

\author{Shen Ai}
\affiliation{Center for Soft Matter Research, Department of Physics, New York University, New York 10003, USA}
\author{Dov Levine}
\affiliation{Department of Physics, Technion-IIT, 32000 Haifa, Israel}
\author{Paul Chaikin}
\affiliation{Center for Soft Matter Research, Department of Physics, New York University, New York 10003, USA}

\date{\today}

\begin{abstract}
We study a dynamical model of repulsive needles with centers fixed on a square lattice. Needles that overlap  are active and receive a repulsive rotation $ \theta \in [0,\epsilon)$; needles with no overlap do not move. This model exhibits an absorbing state transition. Nematic order was observed for long needles. For a range of needle lengths and $\epsilon$, the system self-organizes into a state with checkerboard symmetry. As $\epsilon$ is increased, this phase melts and becomes disordered. Remarkably, we find that this phase boundary exhibits critical exponents identical to those of the equilibrium 2D Ising model.  

\end{abstract}

\maketitle

\begin{bibunit}[apsrev4-2]
Phase transitions in systems out of equilibrium have received much attention in recent years, and continue to provide surprises.  One such surprise was the emergence of 2D crystals in the active phase of disks driven by biased random organization\cite{galliano2023two, guo2026diffusion}, and the emergence of reduced fluctuations at the critical point\cite{PRL_hyper} in absorbing state models\cite{Non-Equilibrium_Book}.  Phase transitions in absorbing state models are between a dynamical active state and a static absorbing phase, with the fraction of active particles $f_a$ being the order parameter. Additional ordering transitions can also occur, either together with the absorbing-active transition or within the active phase\cite{galliano2023two, ghosh2022coupled, guo2026diffusion}.  

In this paper, we describe a system which undergoes an absorbing/active transition as well as an emergent ordering transition from an isotropic to an antiferromagnetic state. The transition shows critical behavior which appears to be identical to the Ising universality class despite the fact that the interacting degrees of freedom are continuous and do not have Ising symmetry.  Rather, it is the mediation of the interaction with the geometry of the lattice that induces this behavior.

\begin{figure}[h!tbp]
    \centering
    \includegraphics[width=1\columnwidth]{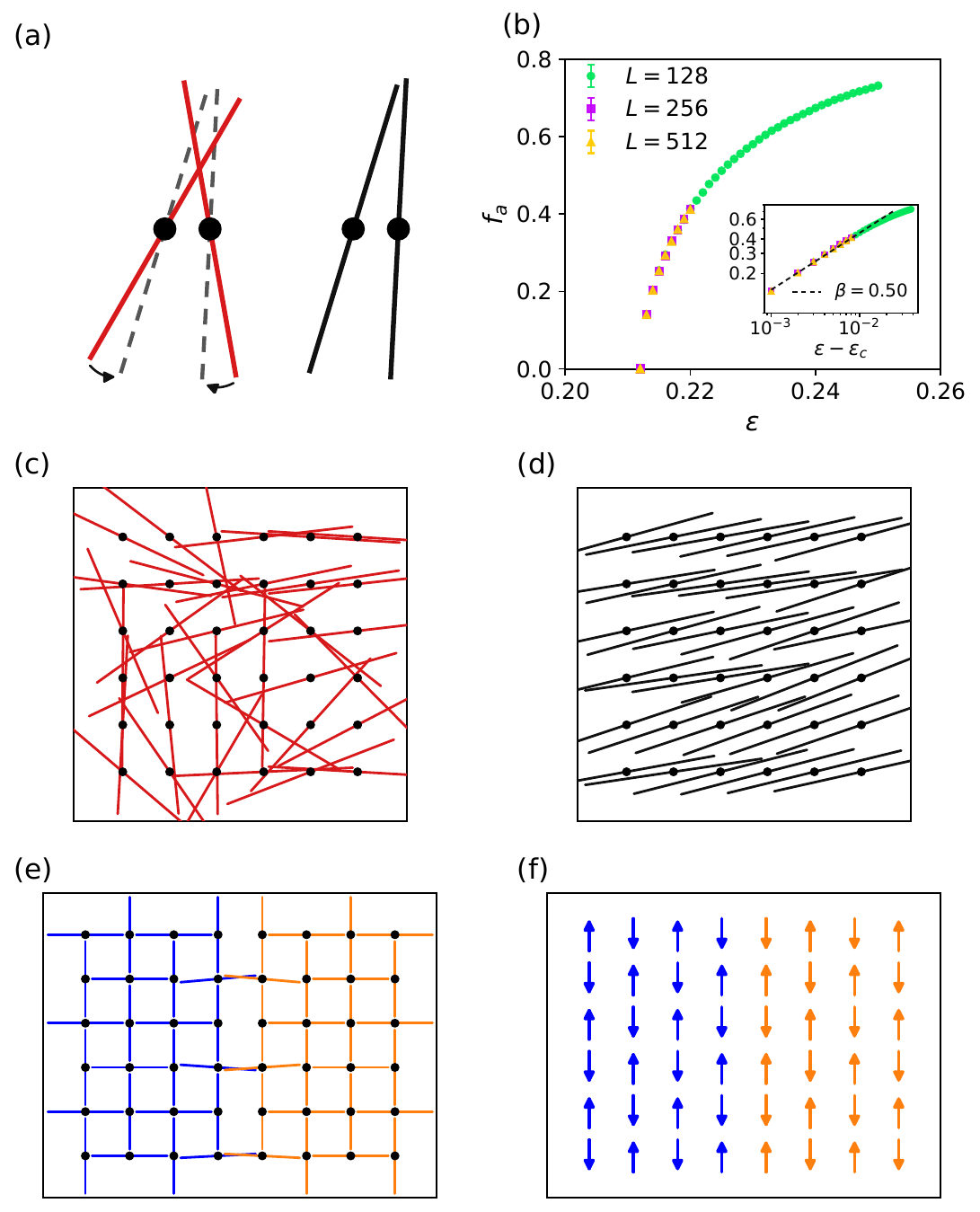}
    \caption{\label{Fig.1}
        (a) Needles with half-length $r=4$, centers fixed on a square lattice with spacing 1 ; BRO kick after which overlapping particles become inactive.
        (b) Absorbing--active transition at $r=2$: fraction of active particles $f_a$ versus kick size $\epsilon$ (in radians); inset log--log plot with $\beta=0.5$ at $\epsilon_c=0.212$.
        (c,d) Representative configurations at $r=2$ colored by activity (red: active, black: inactive, as in (a)):
        (c) $\epsilon=1$, active random phase;
        (d) $\epsilon=0.2$, absorbing phase with nematic order.
        (e,f) Ideal reference states for the checkerboard (CB) phase (e) and antiferromagnetic Ising model (f); blue and orange denote the two phases; the boundary between the two phases corresponds to a phase shift.
        Central horizontal needles in (e) are offset slightly so both remain visible.}
\end{figure}

\begin{figure*}[!t]
    \centering
    \includegraphics[width=\textwidth]{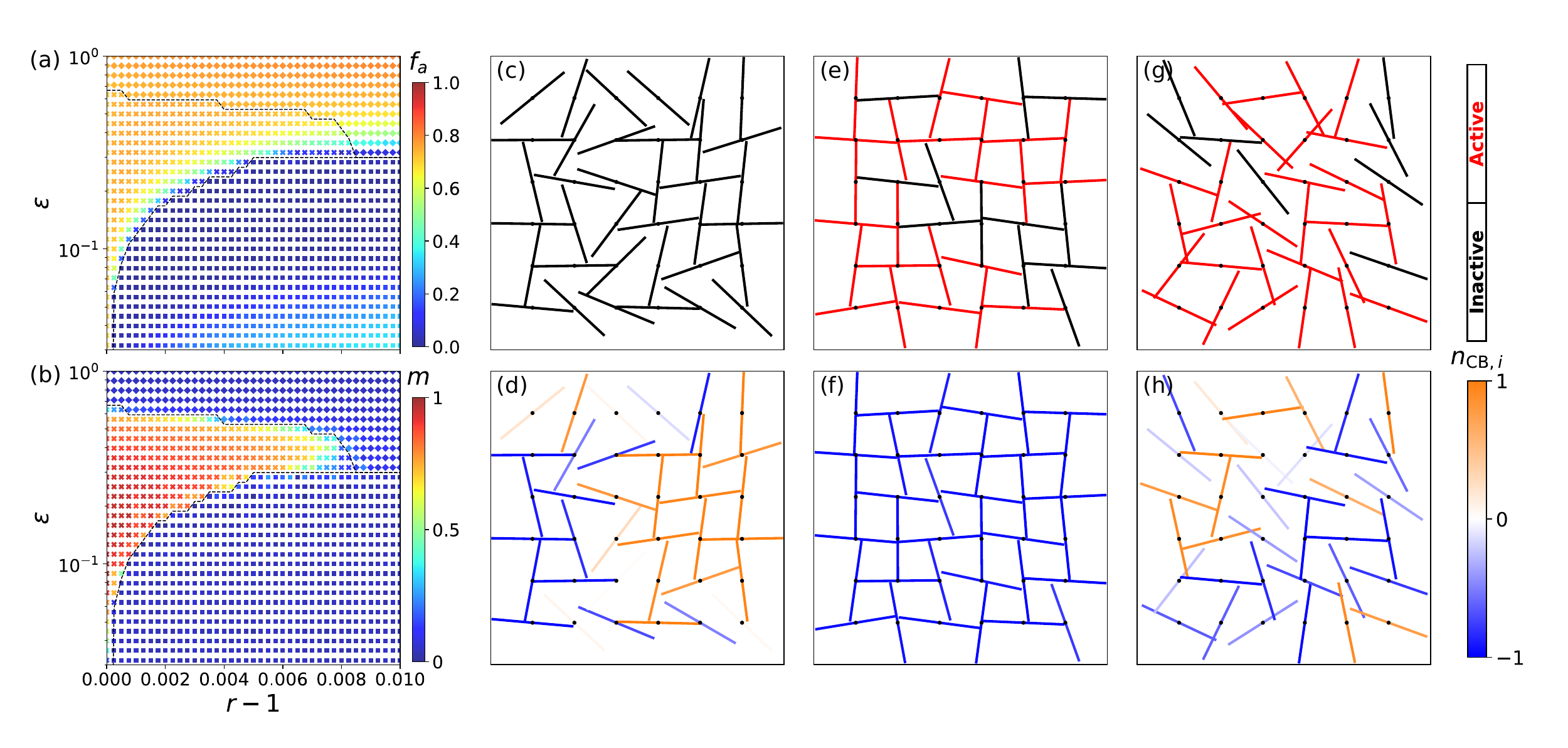}
    \caption{\label{Fig.2}
        Phase diagrams in the $(r-1,\epsilon)$ plane, with markers colored by two order parameters
        (a) the fraction of active particles $f_a$ and
        (b) the CB order parameter $m=|\langle n_{\mathrm{CB},i}\rangle|$.
        Different markers denote the three phases:
        absorbing phase (squares), active CB phase (crosses),
        and active random phase (diamonds).
        Each point is an independent simulation result starting from random initial conditions.
        Dashed lines are guides to the eye for the phase boundaries.
        The lower phase boundary corresponds to a continuous absorbing-to-active transition,
        whereas the upper phase boundary separating the active CB and active random phases
        belongs to the Ising universality class.
        (c--h) Representative configurations in the three phases:
        absorbing phase (c,d) with $r=1.001$ and $\epsilon=0.1$, active CB phase (e,f) with $r=1.001$ and $\epsilon=0.25$, and active random phase (g,h) with $r=1.001$ and $\epsilon=1$.
        Each pair shows the same configuration with two coloring schemes:
        particles are colored by activity in (c,e,g), where red and black indicate active and inactive particles,
        and by the local CB order parameter $n_{\mathrm{CB},i}$ in (d,f,h), where blue and orange denote the two variants. 
        Nematic order, shown for longer needles in Fig.~\ref{Fig.1}(d), does not appear in this range of $r$.
        }
\end{figure*}

The model consists of needles with zero width and length $2r$, whose centers are fixed on a 2D square lattice with unit lattice constant.     The dynamics we employ is reminiscent  of the Biased Random Organization (BRO) \cite{wilken2021random} model, a variant of the Random Organization (RO) \cite{corte} model. A needle is considered active if it crosses (overlaps)  at least one other needle. At each time step, each active needle rotates in the direction that reduces overlap of the two needles; see below. The size of the rotation angle is chosen randomly from a flat distribution between $0$ and $\epsilon$ radians; we call $\epsilon$ the kick size. The dynamics we employ is parallel, where all active needles are updated synchronously at each time step, and  periodic boundary conditions are employed.

To specify this interaction geometrically, we first consider a single overlapping pair of needles, as shown in Fig.~\ref{Fig.1}(a).
We define the pairwise overlap by extending the contact function of ellipses~\cite{perram1985contact, donev2007nonspherical} to the zero-width (needle) limit.
The biased rotation of each needle is then the direction that decreases this overlap.
If a needle overlaps several neighbors, its net rotation direction is a weighted combination of the pairwise contributions.
The geometric construction and mathematical definition are given in the Supplemental Material~\cite{SM}.

This model exhibits an absorbing/active transition characterized by the fraction of active particles $f_a=N_A/N$, where $N_A$ is the number of active needles and $N$ is the total number of needles. Trivial absorbing states (with $f_a=0$) occur when $r<\frac{1}{2}$, since overlapping is impossible.   Similar to the RO and BRO models, the phase diagram of the system depends on two parameters, the kick size $\epsilon$ and the needle length $2r$.  For given $r$, $f_a$ increases as $\epsilon$ increases; this is seen in Fig.~\ref{Fig.1}(b) for $r=2$ where we see a continuous absorbing--active transition with critical point $\epsilon_c \approx 0.212$. The inset log-log plot shows the critical behavior with exponent $\beta=0.5$, which is different from the Manna universality class \cite{Non-Equilibrium_Book,MenonRamaswamy2009}.

\begin{figure*}[!t]
    \centering
    \includegraphics[width=\textwidth]{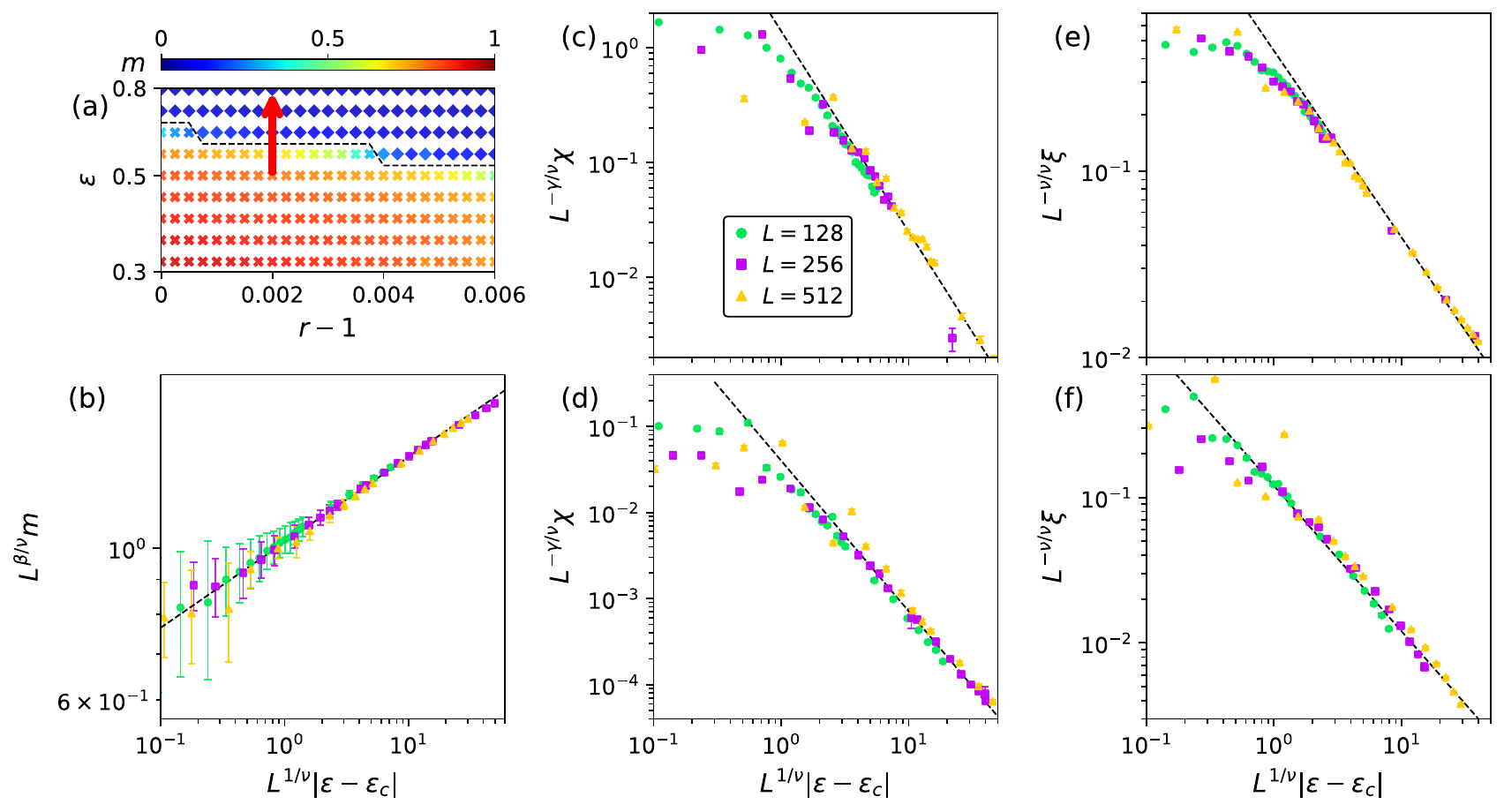}
    \caption{\label{Fig.3}
        Finite-size scaling analysis near the upper phase boundary.
        (a) Phase diagram in the $(r-1,\epsilon)$ plane, colored by $m$; the red arrow indicates the cut at $r-1=0.002$ used for the scaling analysis.
        Results are shown for linear system sizes $L=128$, $256$, and $512$, as indicated by the shared legend.
        All scaling plots use the critical point $\epsilon_c=0.5845$.
        (b) Rescaled CB order parameter $L^{\beta/\nu} m$ on the CB side.
        (c,d) Rescaled susceptibility $L^{-\gamma/\nu}\chi$ on the random side (c) and the CB side (d).
        (e,f) Rescaled correlation length $L^{-1}\xi$ on the random side (e) and the CB side (f).
        Dashed lines indicate the expected two-dimensional Ising critical exponents, $\gamma/\nu=7/4$, $\beta/\nu=1/8$, and $\nu=1$.
        The data collapse and asymptotic power-law behavior are consistent with the 2D Ising universality class.}
\end{figure*}

For long enough needles and large enough $\epsilon$, the system will be active and the needle orientation will be random; this is seen in Fig.~\ref{Fig.1}(c) for $r=2$, $\epsilon=1$. Due to the repulsive nature of the interaction, at smaller $\epsilon$, we might expect needles to align parallel to their neighbors. This indeed occurs, as seen in Fig.~\ref{Fig.1}(d), where the system is in a nematic absorbing state with $r=2$ and $\epsilon \approx 0.21$.  This nematic ordering is not the focus of the present paper.

Surprisingly, when $r$ is slightly larger than the lattice spacing, we observe an active phase where needles align largely perpendicular to their neighbors. We call this self-organized state a checkerboard (CB) phase, as the particles alternate vertical and horizontal states on the 2D square lattice as on a checkerboard. The CB order may be thought of as anti-ferromagnetic.  Fig.~\ref{Fig.1}(e) provides an idealized sketch of a perfect CB state; in the actual system, needles fluctuate around the vertical ($\theta=\pi/2$) or horizontal ($\theta =0$) orientations.  To characterize the CB phase, we define a local CB order parameter on needle $i$ at lattice site $(i_x,i_y)$ and the global order parameter,
\begin{equation}
\begin{aligned}
n_{\mathrm{CB},i} &= (-1)^{i_x+ i_y}\cos 2\theta_{i}\,\\
m &= \bigl|\langle n_{\mathrm{CB}, i}\rangle\bigr|\,
\end{aligned}
\end{equation}
where $\langle \cdot  \rangle$ is an average over both space and time.
The $\cos 2\theta_{i}$ respects the head-tail symmetry of the needles, while the staggered factor reflects the checkerboard pattern, similar to the antiferromagnetic Ising model.

When $\epsilon$ is large enough (such as $\epsilon= 2\pi$), all active needles re-orient randomly and the CB phase  disappears. This suggests  the existence of an active random phase with $m=0$  at large $\epsilon$. On the other hand, when $\epsilon$ is small enough, needles may avoid all the neighbors and the system may organize itself into absorbing state.

These considerations are quantified in the phase diagrams, shown in Fig.~\ref{Fig.2}(a) and (b). In Fig.~\ref{Fig.2}(a), the fraction of active particles $f_a$ is plotted, showing the dynamical transition from an absorbing to an active phase.  Fig.~\ref{Fig.2}(b) shows the ordering transition, where the CB order parameter $m$ is plotted. Taken together, three behaviors are observed: an absorbing phase with $f_a=0$, an {\it active CB phase} with $f_a>0$ and $m\neq 0$, and an {\it active random phase} with $f_a>0$ and $m=0$. Correspondingly, there are two phase boundaries: the lower boundary corresponds to the absorbing-active transition, and the upper boundary separates the active CB and active random phases.

Figs.~\ref{Fig.2}(c)--\ref{Fig.2}(h) show representative  snapshots from each phase.
Each phase is shown in two ways on the same configuration: (c,e,g) color needles by activity (red active, black inactive), and (d,f,h) color the same snapshot by the local CB order parameter (blue and orange for different sign of $n_{\mathrm{CB},i}$).
The absorbing examples (c,d) lie in the lower $f_a=0$ region of the phase diagram. With the random initial conditions used, the system stays disordered. Note here the needles are not long enough to form nematic order.
The active CB phase examples (e,f) show the alternating horizontal and vertical pattern. They also show imperfect checkerboard rather than the perfect alternating pattern sketched in Fig.~\ref{Fig.1}(e).
The active random example (g,h) shows the system is very active while the orientations of the needles are random. 

\begin{figure*}[!ht]
    \centering
    \includegraphics[width=0.9\textwidth]{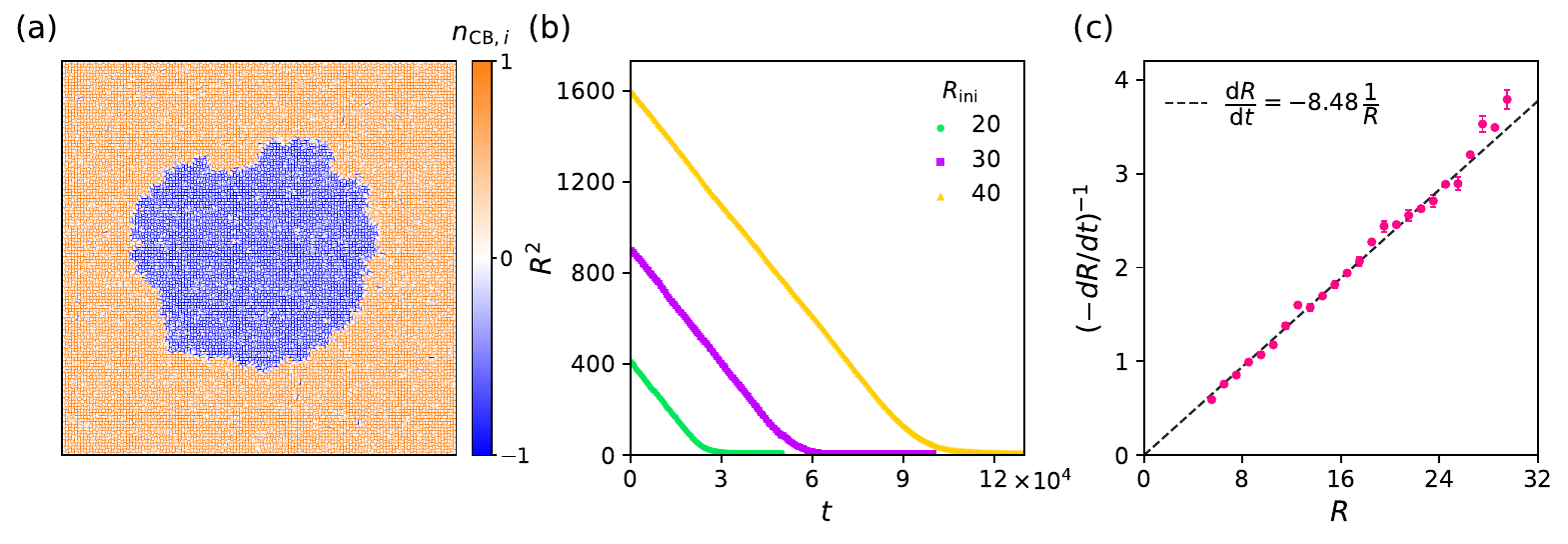}
    \caption{\label{Fig.4}
        (a) Example configuration with a circular center phase ($n_{\mathrm{CB},i}=-1$) surrounded by the outer phase ($n_{\mathrm{CB},i}=+1$); needles are colored by the local CB order parameter $n_{\mathrm{CB},i}$.
        (b) Center-phase size dynamics: independent of the initial center size, the squared effective radius $R^2$ of the center phase decays linearly with time.
        (c) Direct verification of surface-tension behavior: $(-dR/dt)^{-1}$ versus $R$; the dashed line is a fit showing expected coarsening behavior; $\mathrm{d}/\mathrm{d}t$ is evaluated per $10^3$ simulation steps.}
\end{figure*}

The disappearance of the active CB phase for large enough $\epsilon$ is not surprising. What is remarkable, however, is the nature of the  order-disorder transition between the two active phases.  In particular, we find that this phase boundary appears to belong to the Ising universality class, with the order parameter $m$ serving as the magnetization-like variable and $\epsilon$ serving as the temperature-like control parameter.

We performed simulations along $r-1=0.002$ with different $\epsilon$ as indicated in Fig.~\ref{Fig.3}(a) with system sizes $L=128$, $256$, and $512$. Figs.~\ref{Fig.3}(b)--\ref{Fig.3}(f) show  finite-size scaling analysis, with critical point $\epsilon_c=0.5845$.
Fig.~\ref{Fig.3}(b) shows the rescaled CB order parameter $L^{\beta/\nu}m$ on the CB side of the boundary, decaying as $\epsilon$ approaches $\epsilon_c$ from below. Fig.~\ref{Fig.3}(c) and (d) show the rescaled susceptibility 
\footnote{The susceptibility is defined by
$$
    \chi
    =
    \frac{N}{\epsilon}
    \left(
    \left\langle n_{\mathrm{CB}}^2 \right\rangle
    -
    \left\langle n_{\mathrm{CB}} \right\rangle^2
    \right)
$$ where $n_{\mathrm{CB}} = \frac{1}{N}\sum n_{\mathrm{CB}, i}$} 
$L^{-\gamma/\nu}\chi$ on the random side (c) and the CB side (d).
Fig.~\ref{Fig.3}(e) and (f) show the rescaled correlation length $L^{-1}\xi$ on the random side (e) and the CB side (f). The data for all three observables collapse onto common curves, and the dashed lines are guides that are consistent with the two-dimensional Ising exponents $\beta/\nu=1/8$, $\gamma/\nu=7/4$, and $\nu=1$.

The above is true even though our model is fundamentally out of equilibrium: a pair of active needles can become inactive through a kick, but the reverse process is impossible. Even if we coarsen the degrees of freedom into two states: ``vertical" and ``horizontal", our analysis in the Supplemental Material~\cite{SM} shows that entropy production~\cite{Kurchan1998,LebowitzSpohn1999,ro2022model} of the system is non-zero. 
Additionally, the Ising order in our model is emergent in that the microscopic degrees of freedom (the needles) do not have the symmetry of an Ising spin.  This differs from the majority of previous work on nonequilibrium Ising-like systems,
 such as multi-bath coupling systems~\cite{WangLebowitz1988NESSIsing,Marques1989LocallyCompetingTemperatures,Blote1990TwoTemperatures,Tamayo1994TwoTemperatureIsing, Achahbar1996NonequilibriumIsing},
  dynamic rule-based systems~\cite{Oliveira1992MajorityVote,OliveiraMendesSantos1993IsingUniversal,Marro1994NonHamiltonian,MendesSantos1998ShortTimeMajorityVote,VieiraCrokidakis2016TwoNoises,VilelaStanley2018StrongOpinions},
   driven kinetic Ising systems~\cite{Korniss2000DynamicPhaseTransition,BuendiaRikvold2008SoftDynamics,VatanseverFytas2018QuenchedRandomness,Vatansever2024NonantisymmetricField,HalderAcharyya2016StandingWave,HalderAcharyya2017SpinSWaves},
   and the active Ising model at vanishing self-propulsion~\cite{SolonTailleur2015ActiveIsing},
   which employ spin systems with special dynamical rules. 
Although it is clear that it is the influence of the geometry of the lattice of needles that leads to the Ising ordering, this need not be the case, as evidenced by the existence of nematic ordering in portions of the phase space.

From where does the Ising symmetry arise? The square lattice provides a four-fold symmetric ``crystal field", which suggests that the states of the system are degenerate with respect to four-fold rotations. Further, the nematic geometry of the needles reduces the number of degenerate states by two from $\theta$, $\theta + \frac{\pi}{2}$,  $\theta+\pi $, and $\theta + \frac{3\pi}{2}$, being equivalent states to only $\theta$,  and $\theta+\pi $ being equivalent.
However, unlike Ising spins, which are discrete and only take on values ``up'', $\frac{\pi}{2}$, or ``down'', $\frac{3\pi}{2}$, the needles can take on continuous values with two sets of degenerate states: horizontal $-\frac{\pi}{4}<\theta<\frac{\pi}{4}$ and vertical $\frac{\pi}{4}<\theta<\frac{3\pi}{4}$.

However, the symmetry similarity between our needle system and Ising model guarantees neither the formation of an active CB phase nor the existence of the transition. This two-fold symmetry argument holds for any $r>0.5$, yet we have shown that  nematic behavior obtains at larger $r$, rather than a CB phase. In fact, as shown in Fig.~\ref{Fig.2}(a) and (b), the CB phase only exists within a very small range $0<r-1<0.006$. 
Existing work \cite{PhysRevB.16.1217} on the relationship between symmetry and universality class differs from our study in that our system is out of equilibrium, and the Ising symmetry is emergent.

Finally, we may ask what happens to a bubble of a ``metastable" phase.  To this end, we study the shrinkage of a minority CB domain embedded in the opposite phase.
Fig.~\ref{Fig.4}(a) shows a circular center region with $n_{\mathrm{CB},i}=-1$ surrounded by the complementary phase with $n_{\mathrm{CB},i}=+1$.
Driven by this interface, the bubble shrinks in time.
For a curved interface, and assuming surface tension, the expected coarsening law is 
\begin{equation}
\frac{\mathrm{d}R}{\mathrm{d}t} \propto -\frac{1}{R}
\label{coarsening}
\end{equation}
 where $R$ is the local radius of curvature\cite{ALLEN19791085}

For a circular geometry, we expect $R(t)^2 = R(0)^2 - K t$. Fig.~\ref{Fig.4}(b) shows $R^2$ versus time for several initial center sizes, and all curves show perfect linear decay with time, consistent with Equation \ref{coarsening}. Note that the decay behavior of individual runs is subject to fluctuations and our results here are averaged over hundreds of realizations.
A direct test of the coarsening law is shown in Fig.~\ref{Fig.4}(c), where $(-\mathrm{d}R/\mathrm{d}t)^{-1}$ is plotted against $R$; the dashed line is a straight line with zero intercept, showing the data follows Equation \ref{coarsening}.

In equilibrium Ising models or in the alloy mixture models \cite{ALLEN19791085}, such interface motion or surface tension behavior is tied to a free energy. 
However, here the microscopic dynamics is purely kinetic: needles evolve by steric repulsion rules, and there is no obvious free energy functional.

In conclusion, we have shown that needles with BRO-like nonequilibrium dynamics on a square lattice can self-organize into an antiferromagnetic-like phase in which each needle tends to align perpendicular to its neighbors. The phase diagram exhibits two distinct boundaries: a continuous absorbing-active transition at the lower boundary and an order-disorder transition at the upper boundary. The order-disorder boundary appears to have Ising universality.  Our system provides a novel type of nonequilibrium model in which Ising-like behavior emerges due to interaction between the continuous degrees of freedom and the geometry of the lattice. 
Along with previous work, our studies show that absorbing state systems can evolve components into complex ordered architectures, especially in their active states. Whether we can program or teach them dynamical rules to form useful moving or static structures is an interesting challenge.

We thank Alexander Grosberg and Mathias Casiulis for useful interactions. DL thanks the Israel Science Foundation for grant 2083/23 and the NSF-BSF (Grant 2023669). PC and DL thank NSF-BSF grant 2414721. This work was supported in part through the NYU IT High Performance Computing resources, services, and staff expertise.

\putbib[\jobname Notes,BRO_needle]
\end{bibunit}

\clearpage
\onecolumngrid

\setcounter{topnumber}{2}
\setcounter{bottomnumber}{1}
\setcounter{totalnumber}{3}
\renewcommand{\topfraction}{0.92}
\renewcommand{\bottomfraction}{0.80}
\renewcommand{\textfraction}{0.08}
\renewcommand{\floatpagefraction}{0.78}
\setlength{\textfloatsep}{14pt plus 3pt minus 4pt}
\setlength{\floatsep}{12pt plus 3pt minus 3pt}

\renewcommand{\thefigure}{S\arabic{figure}}
\renewcommand{\thetable}{S\arabic{table}}
\renewcommand{\theequation}{S\arabic{equation}}
\setcounter{figure}{0}
\setcounter{table}{0}
\setcounter{equation}{0}
\setcounter{section}{0}
\setcounter{secnumdepth}{1}
\makeatletter
\renewcommand{\thesection}{\Alph{section}}
\renewcommand{\@seccntformat}[1]{\csname the#1\endcsname.\hspace{0.6em}}
\renewcommand{\section}{\@startsection{section}{1}{\z@}%
  {2.4ex plus 0.8ex minus 0.3ex}%
  {1.2ex plus 0.2ex}%
  {\normalfont\large\bfseries}}
\makeatother

\begin{center}
{\large\bfseries Supplemental Material for ``An Absorbing State Model with an Emergent Ising Transition''\par}
\vspace{0.75em}
{\normalsize Shen Ai,$^{1}$ Dov Levine,$^{2}$ and Paul Chaikin$^{1}$\par}
\vspace{0.35em}
{\small $^{1}$Center for Soft Matter Research, Department of Physics, New York University, New York 10003, USA\\
$^{2}$Department of Physics, Technion-IIT, 32000 Haifa, Israel\par}
\end{center}
\vspace{1.2em}

\begin{bibunit}[apsrev4-2]
\section{Model Details}
\label{sec:model-details}

\begin{figure}[htbp]
    \centering
    \includegraphics[width=0.96\textwidth]{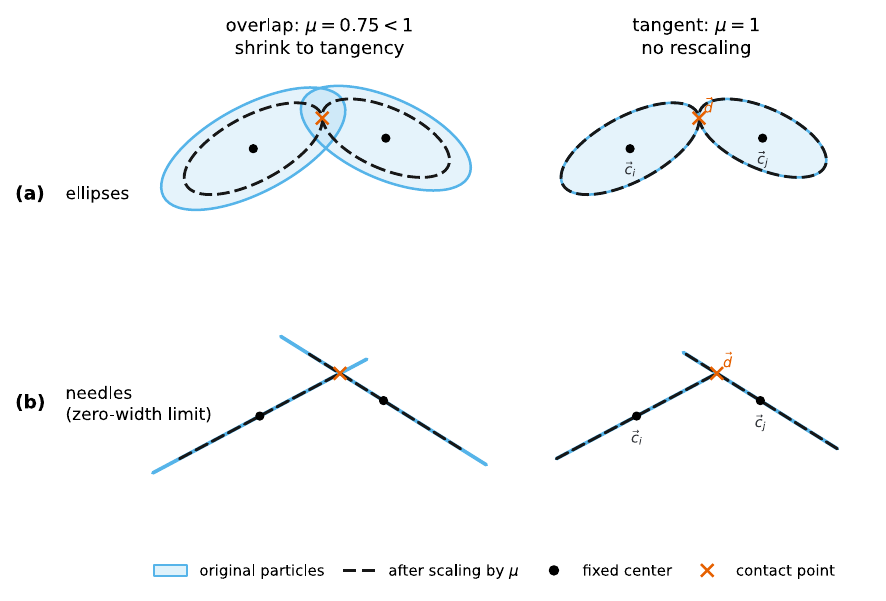}
    \caption{Common-scaling construction for the contact function, shown for the overlapping ($\mu<1$) and tangency ($\mu=1$) cases. Each row shows an overlapping pair (left) and a tangent pair (right). (a)~Two ellipses with fixed centers and orientations are scaled uniformly by $\mu$ until their boundaries are tangent. (b)~The analogous construction in the zero-width limit gives two needles. Colored particles show the original configurations, gray dashed outlines show the configurations after scaling by $\mu$, black points mark the fixed centers $\vec{c}_i$ and $\vec{c}_j$, and red crosses mark the contact points.}
    \label{fig:si-a1}
\end{figure}

In this section, we introduce details of our simulation model. The pairwise overlap $\mu_{ij}$ is the zero-width limit of the contact function
introduced by Perram and Wertheim for ellipses and ellipsoids~\cite{perram1985contact, donev2007nonspherical}.
For two ellipses with fixed centers and orientations, let $\mu$ be the common
homothetic scale factor that brings their boundaries into tangential contact.
Then $\mu<1$ indicates overlap, $\mu=1$ tangency, and $\mu>1$ separation.
Taking the minor axes to zero gives the corresponding contact construction for needles.
Fig.~\ref{fig:si-a1} illustrates only the overlapping ($\mu<1$) and tangency ($\mu=1$) cases.
The general contact-function construction also permits $\mu>1$, corresponding to separated particles; that case is not shown because non-overlapping needles no longer interact in our model.

\FloatBarrier

Each needle $i$ has half-length $r$ and a fixed center $\vec{c}_i$ at lattice
index $(i_x,i_y)$. For a pair of needles $i$ and $j$, let $\vec{d}$ denote the
intersection point of the infinite lines containing the needles, obtained by
extending the finite needles when necessary. We define the pair overlap
variable $\mu_{ij}$ as the common expansion or contraction factor required for
the two needles to barely touch:
\begin{equation}
    \mu_{ij}
    = \frac{\max\!\left(
        \left\lVert\vec{c}_i-\vec{d}\right\rVert,
        \left\lVert\vec{c}_j-\vec{d}\right\rVert
      \right)}{r}.
    \label{eq:si-overlap-metric}
\end{equation}
Thus, $\mu_{ij}<1$ corresponds to overlapping needles, $\mu_{ij}=1$ to needles
that barely touch, and $\mu_{ij}>1$ to non-overlapping needles. For parallel
needles, the intersection point recedes to infinity and
$\mu_{ij}\rightarrow\infty$. The dimensionless quantity $\mu_{ij}$ is therefore a
scale-invariant measure of the pairwise ``distance'' between two needles.

For an overlapping pair of needles, the biased direction of motion that tends to separate them should increase their ``distance'' $\mu_{ij}$.
Therefore, the signed direction in which needle $i$ rotates in response to an overlapping
neighbor $j$ is defined as
\begin{equation}
    \hat{\theta}_{i,j}
    = \sgn\!\left(\frac{\partial \mu_{ij}}{\partial\theta_i}\right).
    \label{eq:si-pair-rotation-direction}
\end{equation}
If needle $i$ has multiple overlapping neighbors, denoted by the set
$\mathcal{O}_i$, its net rotation direction follows the weighted sum of their
contributions:
\begin{equation}
    \hat{\theta}_i
    =
    \sgn\!\left[
        \sum_{j\in\mathcal{O}_i}
        \left|
            \frac{\partial \mu_{ij}}{\partial \theta_i}
        \right|
        \hat{\theta}_{i,j}
    \right]
    =
    \sgn\!\left[
        \sum_{j\in\mathcal{O}_i}
        \frac{\partial \mu_{ij}}{\partial \theta_i}
    \right].
    \label{eq:si-net-rotation-direction}
\end{equation}

\section{Orientation Distributions and Boundary Geometry}
\label{sec:orientation}

The needle orientations are continuous and have head-tail (nematic) symmetry, so
$\theta$ and $\theta+\pi$ describe the same physical state. 
$\theta=0$ corresponds to the horizontal orientation, and $\theta=\pi/2$ corresponds to the vertical orientation.
We use the one-particle distribution $P(\theta)$ to characterize the local orientational behavior.

\begin{figure}[htbp]
    \centering
    \includegraphics[width=0.6\textwidth]{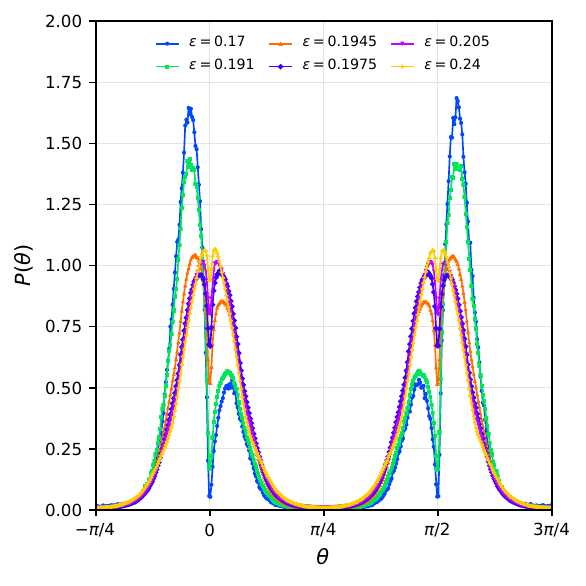}
    \caption{Orientation probability density $P(\theta)$ versus orientation angle $\theta$ near the absorbing-active boundary with $r=1.002$ and system size $N=512^2$. When the system is active, $P(\theta)$ is symmetric around $\theta=0$ and $\theta=\pi/2$. When the system is absorbing, $P(\theta)$ is asymmetric around $\theta=0$ and $\theta=\pi/2$ but stays symmetric around $\theta=\pi/4$.}
    \label{fig:si-b1}
\end{figure}

As shown in Fig.~\ref{fig:si-b1}, near the absorbing-active phase boundary in Fig.~2 of the main text, the symmetry of $P(\theta)$ changes across the phase boundary.
In the active state, $P(\theta)$ is symmetric around $\theta=0$ and $\theta=\pi/2$, suggesting that the deviations from horizontal or vertical are symmetric.
When the system is absorbing, $P(\theta)$ becomes asymmetric around $\theta=0$ and $\theta=\pi/2$ but stays symmetric around $\theta=\pi/4$.
Although the needles still populate the vertical and horizontal orientations, the symmetry of the deviation differs between absorbing and active states.

\FloatBarrier

\begin{figure}[htbp]
    \centering
    \includegraphics[width=0.65\textwidth]{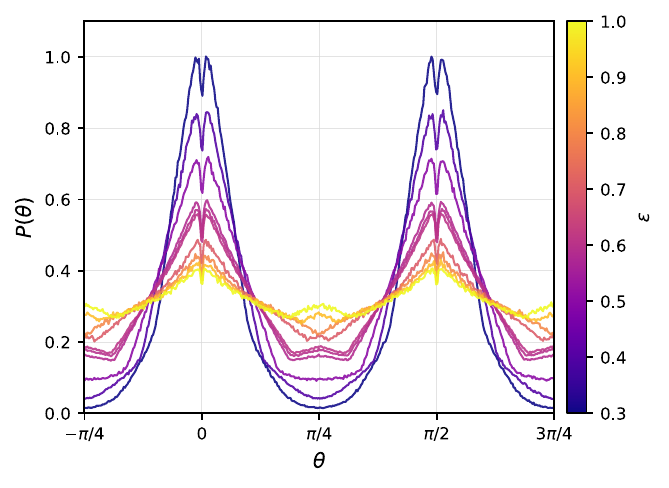}
    \caption{Orientation probability density $P(\theta)$ versus orientation angle $\theta$ for kick sizes from $\epsilon=0.3$ (CB phase) to $\epsilon=1.0$ (active random phase) at $r=1.002$ and system size $N=512^2$; the color of the curves denotes $\epsilon$. As $\epsilon$ increases, $P(\theta)$ broadens. Several curves with similar $\epsilon$ around the Ising transition are shown, and no obvious distinction is observed. Unlike the absorbing-active transition, the Ising transition is not related to $P(\theta)$, but to the spatial correlation of needle orientations.}
    \label{fig:si-b2}
\end{figure}

Fig.~\ref{fig:si-b2} shows $P(\theta)$ for a larger range of $\epsilon$ from $\epsilon=0.3$ to $\epsilon=1.0$ (always in the active phase). $P(\theta)$ peaks at $\theta=0$ and $\theta=\pi/2$, meaning the system prefers to align horizontally or vertically, even at large $\epsilon$.
As one would expect, $P(\theta)$ broadens as $\epsilon$ increases. The preference for horizontal or vertical orientation becomes less pronounced yet exists as $\epsilon$ increases.
Multiple curves with similar $\epsilon$ around the Ising transition $\epsilon_c=0.5845$ are shown, and no obvious distinction is observed. This confirms that the Ising transition from the checkerboard (CB) phase to the active random phase corresponds to the disappearance of the spatial correlation of needle orientations, but not to a change in $P(\theta)$ itself.

\FloatBarrier

\begin{figure}[htbp]
    \centering
    \includegraphics[width=0.65\textwidth]{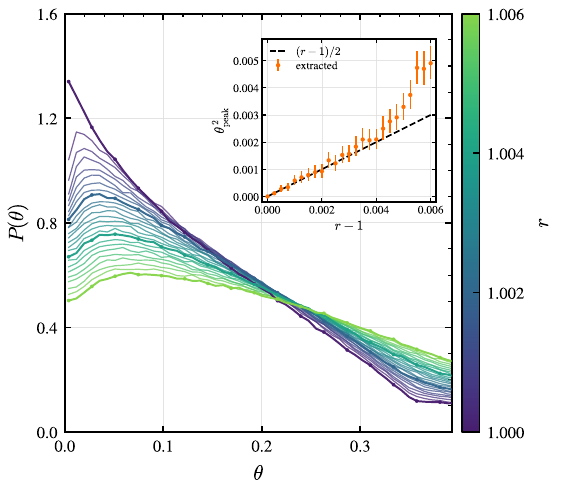}
    \caption{Folded orientation probability density $P(\theta)$ versus orientation angle $\theta$ for 25 needle half-lengths $r=1.000\sim1.006$ at fixed kick size $\epsilon=0.355$ and system size $N=256^2$; color denotes $r$. As $r$ increases, $P(\theta)$ develops a maximum at a nonzero angle. Inset: squared peak position $\theta_{\mathrm{peak}}^2$ versus $r-1$. The data points in the CB phase follow the dashed line $\theta_{\mathrm{peak}}^2=(r-1)/2$, and the deviation from the dashed line occurs around the phase boundary.}
    \label{fig:si-b3}
\end{figure}

Another observation, as shown in Fig.~\ref{fig:si-b1}, is that $P(\theta)$ peaks near, but also dips at, $\theta=0$ and $\theta=\pi/2$.
The origin of the dips is $r$, the half-length of the needles.
Fig.~\ref{fig:si-b3} shows the folded orientation probability density $P(\theta)$ for 25 needle half-lengths $r=1.000\sim1.006$ at fixed kick size $\epsilon=0.355$.
When $r=1$, i.e., the half-length of the needles is equal to the lattice constant, the dips disappear and $P(\theta)$ peaks at $\theta=0$ (and $\theta=\pi/2$, though not shown).
As $r$ increases, $P(\theta)$ develops a maximum at a nonzero $\theta$ and the dips become more pronounced.
This observation supports the idea that the dips arise from the needle half-length.
Further, as the inset shows, the peak position follows $\theta_{\mathrm{peak}}^2 \simeq (r-1)/2$ in the CB phase and only deviates from the dashed line around the phase boundary.

There is a geometric explanation behind this relationship: imagine that one horizontal needle has a perfectly vertical neighbor. For the needle to avoid its neighbor, it needs to deviate by a small angle such that $r\cos(\theta)=1$. Since $r-1$ and $\theta$ are both small, by Taylor expansion, we have $\theta^2 \propto (r-1)$, which is consistent with the scaling behavior in the inset.
This observation and geometric explanation further confirm that the origin of the dips is the needle half-length.

\FloatBarrier

\section{Active Nematic Configuration}
\label{sec:active-nematic}

\begin{figure}[htbp]
    \centering
    \includegraphics[width=0.7\textwidth]{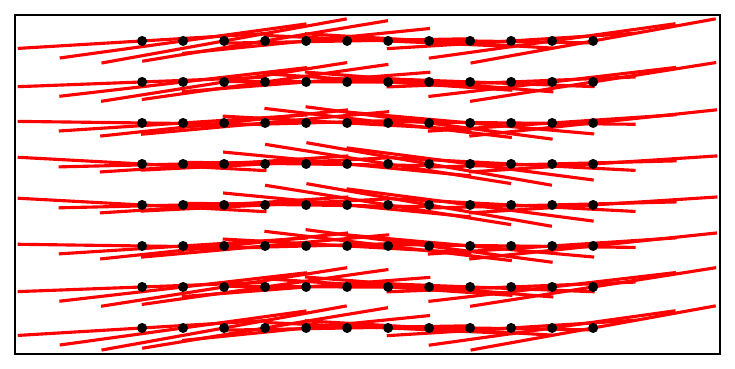}
    \caption{Active nematic configuration with $r=3$ and $\epsilon=0.5$. Red lines show active needles, and black dots mark their fixed centers. The common alignment of the needles shows the nematic order.}
    \label{fig:si-c1}
\end{figure}

In Fig.~1 of the main text, we showed an absorbing nematic configuration with $r=2$. 
However, nematic order does not exist only in the absorbing state. 
As shown in Fig.~\ref{fig:si-c1}, we can also observe an active nematic configuration with $r=3$ and $\epsilon=0.5$.

\FloatBarrier

\section{Entropy Production and Irreversibility}
\label{sec:entropy-production}

The Ising critical behavior of the model may make one wonder whether the system is in equilibrium, at least near the upper phase boundary, and whether the control parameter $\epsilon$ may play a role similar to temperature. The physical effect of $\epsilon$, which increases the fluctuations of the system when it is larger, also suggests its similarity to temperature. Therefore, we first explore the idea of an ``effective temperature.'' By defining various types of energy, from pairwise to many-body, and a temperature based on $\epsilon$, we try to reproduce the Boltzmann distribution from the simulation results, which, if successful, would suggest that the system is in equilibrium and can be simulated by thermodynamic sampling. However, none of these attempts succeeded, which leads to our further investigation of the entropy production.

Microscopically, only overlapping needles are updated, and their rotations are biased toward reducing
overlap. This dynamics is fundamentally irreversible: a pair of active overlapping needles may become non-overlapping and inactive after a kick, but a pair of inactive needles will not move.
Although the model is fundamentally out of equilibrium, the emergent system with coarse-grained degrees of freedom may behave as if in equilibrium. 
We test this idea by coarse-graining the continuous orientations into the two states
relevant to CB order.

\begin{figure}[htbp]
    \centering
    \includegraphics[width=0.9\textwidth]{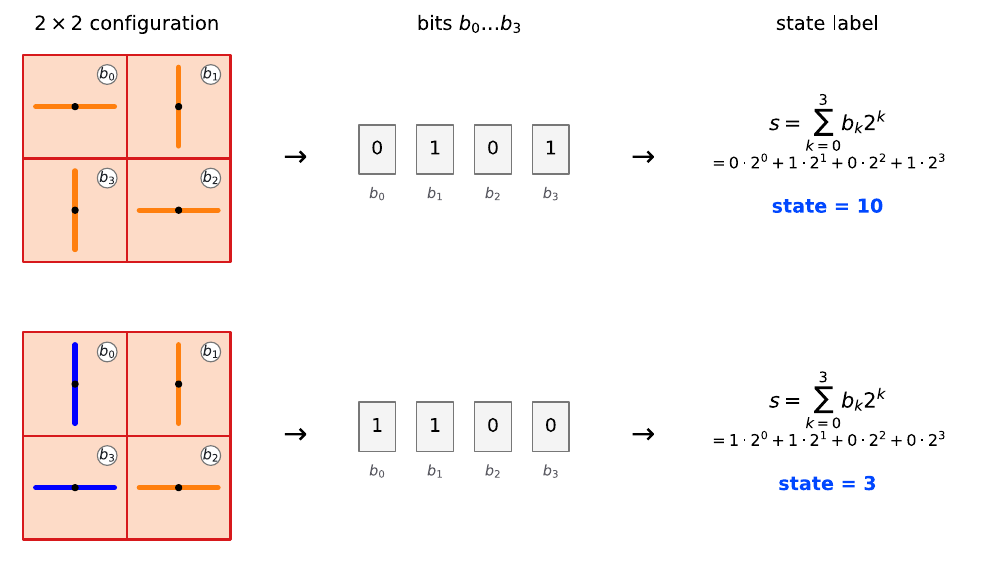}
    \caption{Binary-state construction for a $2\times2$ patch. The sites are ordered clockwise from the upper-left site as $b_0,\ldots,b_3$. Two needle configurations are converted to the bit strings $0101$ and $1100$, corresponding to states 10 and 3 under $s=\sum_{k=0}^{3}b_k2^k$. Needle colors follow the signed CB-order convention used in Fig.~2 of the main text.}
    \label{fig:si-d1}
\end{figure}

For each needle, we define a binary orientation variable
\begin{equation}
    b_i(t) =
    \begin{cases}
        0, & \cos(2\theta_i(t)) \geq 0,\\
        1, & \cos(2\theta_i(t)) < 0,
    \end{cases}
    \label{eq:si-binary-state}
\end{equation}
corresponding to the horizontal and vertical orientation sectors,
respectively. For the $2\times2$ patch, the four sites are ordered clockwise
from the upper-left site, and the instantaneous patch state is the integer
\begin{equation}
    s(t)=\sum_{k=0}^{3} b_k(t)2^k.
    \label{eq:si-patch-state}
\end{equation}
Fig.~\ref{fig:si-d1} shows this mapping for two configurations, which give
states 10 and 3. 

\FloatBarrier

We then measure the entropy production~\cite{Kurchan1998,LebowitzSpohn1999} from the time series of the patch states by cross-parsing lossless compression~\cite{ro2022model}. A sequence of total length $2h$ is divided into a

reference half $S_1$ and a second half $S_2$. If $C(X\vert Y)$ denotes the
cross-parsing complexity of $X$ using $Y$ as the reference sequence, the
Kullback--Leibler estimator used to measure entropy production is
\begin{equation}
    D_{\mathrm{KLD}}
    = \left[C(S_2^{\mathrm{R}}\vert S_1)-C(S_2\vert S_1)\right]
      \frac{\log_2 h}{h},
    \label{eq:si-kld}
\end{equation}
where $S_2^{\mathrm{R}}$ is the time-reversed second half. A time-reversible
sequence gives $D_{\mathrm{KLD}}=0$ in the long-sequence limit,
whereas a positive value means that the reversed sequence is less compatible
with the forward-time reference. We therefore use $D_{\mathrm{KLD}}$ as a
local estimator of the entropy production; small negative values can occur from
finite-sample fluctuations around zero.

\begin{figure}[htbp]
    \centering
    \includegraphics[width=0.80\textwidth]{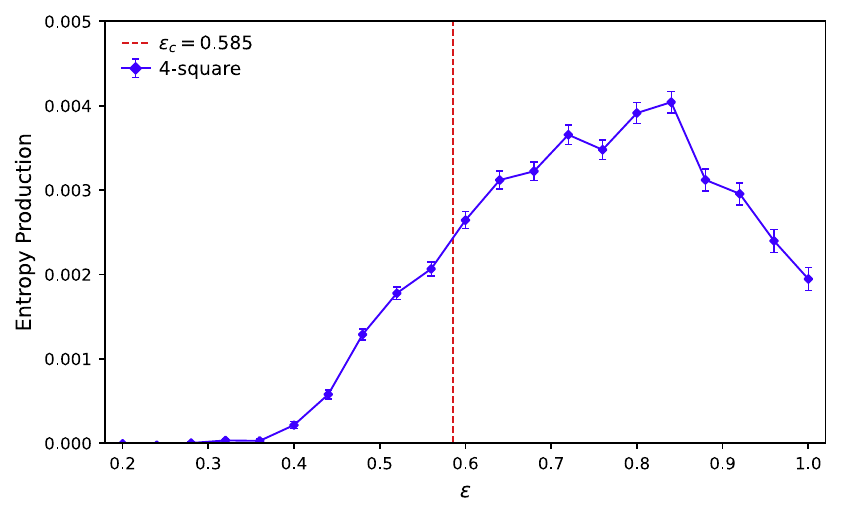}
    \caption{Entropy production versus kick size $\epsilon$ for the $2\times2$ patch. The vertical dashed line marks the CB-order transition at $\epsilon_c=0.5845$.}
    \label{fig:si-d2}
\end{figure}

As shown in Fig.~\ref{fig:si-d2}, entropy production for the $2\times2$ patch, measured with this estimator, becomes nonzero in the active regime
and remains nonzero on both sides of the CB-order transition at
$\epsilon_c=0.5845$. The result clearly shows that the system with coarse-grained degrees of freedom is out of equilibrium. It also confirms that neither an effective temperature nor an energy can be constructed, since the system is fundamentally out of equilibrium.

\FloatBarrier

\putbib[BRO_needle]
\end{bibunit}


\begin{thebibliography}{35}%
\makeatletter
\providecommand \@ifxundefined [1]{%
 \@ifx{#1\undefined}
}%
\providecommand \@ifnum [1]{%
 \ifnum #1\expandafter \@firstoftwo
 \else \expandafter \@secondoftwo
 \fi
}%
\providecommand \@ifx [1]{%
 \ifx #1\expandafter \@firstoftwo
 \else \expandafter \@secondoftwo
 \fi
}%
\providecommand \natexlab [1]{#1}%
\providecommand \enquote  [1]{``#1''}%
\providecommand \bibnamefont  [1]{#1}%
\providecommand \bibfnamefont [1]{#1}%
\providecommand \citenamefont [1]{#1}%
\providecommand \href@noop [0]{\@secondoftwo}%
\providecommand \href [0]{\begingroup \@sanitize@url \@href}%
\providecommand \@href[1]{\@@startlink{#1}\@@href}%
\providecommand \@@href[1]{\endgroup#1\@@endlink}%
\providecommand \@sanitize@url [0]{\catcode `\\12\catcode `\$12\catcode
  `\&12\catcode `\#12\catcode `\^12\catcode `\_12\catcode `\%12\relax}%
\providecommand \@@startlink[1]{}%
\providecommand \@@endlink[0]{}%
\providecommand \url  [0]{\begingroup\@sanitize@url \@url }%
\providecommand \@url [1]{\endgroup\@href {#1}{\urlprefix }}%
\providecommand \urlprefix  [0]{URL }%
\providecommand \Eprint [0]{\href }%
\providecommand \doibase [0]{https://doi.org/}%
\providecommand \selectlanguage [0]{\@gobble}%
\providecommand \bibinfo  [0]{\@secondoftwo}%
\providecommand \bibfield  [0]{\@secondoftwo}%
\providecommand \translation [1]{[#1]}%
\providecommand \BibitemOpen [0]{}%
\providecommand \bibitemStop [0]{}%
\providecommand \bibitemNoStop [0]{.\EOS\space}%
\providecommand \EOS [0]{\spacefactor3000\relax}%
\providecommand \BibitemShut  [1]{\csname bibitem#1\endcsname}%
\let\auto@bib@innerbib\@empty
\bibitem [{\citenamefont {Galliano}\ \emph {et~al.}(2023)\citenamefont
  {Galliano}, \citenamefont {Cates},\ and\ \citenamefont
  {Berthier}}]{galliano2023two}%
  \BibitemOpen
  \bibfield  {author} {\bibinfo {author} {\bibfnamefont {L.}~\bibnamefont
  {Galliano}}, \bibinfo {author} {\bibfnamefont {M.~E.}\ \bibnamefont
  {Cates}},\ and\ \bibinfo {author} {\bibfnamefont {L.}~\bibnamefont
  {Berthier}},\ }\href {https://doi.org/10.1103/PhysRevLett.131.047101}
  {\bibfield  {journal} {\bibinfo  {journal} {Phys. Rev. Lett.}\ }\textbf
  {\bibinfo {volume} {131}},\ \bibinfo {pages} {047101} (\bibinfo {year}
  {2023})}\BibitemShut {NoStop}%
\bibitem [{\citenamefont {Guo}\ \emph {et~al.}(2026)\citenamefont {Guo},
  \citenamefont {Wilken}, \citenamefont {Levine},\ and\ \citenamefont
  {Chaikin}}]{guo2026diffusion}%
  \BibitemOpen
  \bibfield  {author} {\bibinfo {author} {\bibfnamefont {A.~Z.}\ \bibnamefont
  {Guo}}, \bibinfo {author} {\bibfnamefont {S.}~\bibnamefont {Wilken}},
  \bibinfo {author} {\bibfnamefont {D.}~\bibnamefont {Levine}},\ and\ \bibinfo
  {author} {\bibfnamefont {P.~M.}\ \bibnamefont {Chaikin}},\ }\href
  {https://doi.org/10.1103/qqhm-98vn} {\bibfield  {journal} {\bibinfo
  {journal} {Phys. Rev. E}\ }\textbf {\bibinfo {volume} {113}},\ \bibinfo
  {pages} {044108} (\bibinfo {year} {2026})}\BibitemShut {NoStop}%
\bibitem [{\citenamefont {Hexner}\ and\ \citenamefont
  {Levine}(2015)}]{PRL_hyper}%
  \BibitemOpen
  \bibfield  {author} {\bibinfo {author} {\bibfnamefont {D.}~\bibnamefont
  {Hexner}}\ and\ \bibinfo {author} {\bibfnamefont {D.}~\bibnamefont
  {Levine}},\ }\href@noop {} {\bibfield  {journal} {\bibinfo  {journal}
  {Physical Review Letters}\ }\textbf {\bibinfo {volume} {114}},\ \bibinfo
  {pages} {110602} (\bibinfo {year} {2015})}\BibitemShut {NoStop}%
\bibitem [{\citenamefont {Henkel}\ \emph {et~al.}(2008)\citenamefont {Henkel},
  \citenamefont {Hinrichsen},\ and\ \citenamefont
  {L{\"u}beck}}]{Non-Equilibrium_Book}%
  \BibitemOpen
  \bibfield  {author} {\bibinfo {author} {\bibfnamefont {M.}~\bibnamefont
  {Henkel}}, \bibinfo {author} {\bibfnamefont {H.}~\bibnamefont {Hinrichsen}},\
  and\ \bibinfo {author} {\bibfnamefont {S.}~\bibnamefont {L{\"u}beck}},\
  }\href@noop {} {\emph {\bibinfo {title} {Non-Equilibrium Phase Transitions -
  Volume 1: Absorbing Phase Transitions}}}\ (\bibinfo  {publisher} {Springer},\
  \bibinfo {year} {2008})\BibitemShut {NoStop}%
\bibitem [{\citenamefont {Ghosh}\ \emph {et~al.}(2022)\citenamefont {Ghosh},
  \citenamefont {Radhakrishnan}, \citenamefont {Chaikin}, \citenamefont
  {Levine},\ and\ \citenamefont {Ghosh}}]{ghosh2022coupled}%
  \BibitemOpen
  \bibfield  {author} {\bibinfo {author} {\bibfnamefont {A.}~\bibnamefont
  {Ghosh}}, \bibinfo {author} {\bibfnamefont {J.}~\bibnamefont
  {Radhakrishnan}}, \bibinfo {author} {\bibfnamefont {P.~M.}\ \bibnamefont
  {Chaikin}}, \bibinfo {author} {\bibfnamefont {D.}~\bibnamefont {Levine}},\
  and\ \bibinfo {author} {\bibfnamefont {S.}~\bibnamefont {Ghosh}},\
  }\href@noop {} {\bibfield  {journal} {\bibinfo  {journal} {Physical Review
  Letters}\ }\textbf {\bibinfo {volume} {129}},\ \bibinfo {pages} {188002}
  (\bibinfo {year} {2022})}\BibitemShut {NoStop}%
\bibitem [{\citenamefont {Wilken}\ \emph {et~al.}(2021)\citenamefont {Wilken},
  \citenamefont {Guerra}, \citenamefont {Levine},\ and\ \citenamefont
  {Chaikin}}]{wilken2021random}%
  \BibitemOpen
  \bibfield  {author} {\bibinfo {author} {\bibfnamefont {S.}~\bibnamefont
  {Wilken}}, \bibinfo {author} {\bibfnamefont {R.~E.}\ \bibnamefont {Guerra}},
  \bibinfo {author} {\bibfnamefont {D.}~\bibnamefont {Levine}},\ and\ \bibinfo
  {author} {\bibfnamefont {P.~M.}\ \bibnamefont {Chaikin}},\ }\href@noop {}
  {\bibfield  {journal} {\bibinfo  {journal} {Physical review letters}\
  }\textbf {\bibinfo {volume} {127}},\ \bibinfo {pages} {038002} (\bibinfo
  {year} {2021})}\BibitemShut {NoStop}%
\bibitem [{\citenamefont {Cort{\'e}}\ \emph {et~al.}(2008)\citenamefont
  {Cort{\'e}}, \citenamefont {Chaikin}, \citenamefont {Gollub},\ and\
  \citenamefont {Pine}}]{corte}%
  \BibitemOpen
  \bibfield  {author} {\bibinfo {author} {\bibfnamefont {L.}~\bibnamefont
  {Cort{\'e}}}, \bibinfo {author} {\bibfnamefont {P.~M.}\ \bibnamefont
  {Chaikin}}, \bibinfo {author} {\bibfnamefont {J.~P.}\ \bibnamefont
  {Gollub}},\ and\ \bibinfo {author} {\bibfnamefont {D.~J.}\ \bibnamefont
  {Pine}},\ }\href@noop {} {\bibfield  {journal} {\bibinfo  {journal} {Nature
  Physics}\ ,\ \bibinfo {pages} {420}} (\bibinfo {year} {2008})}\BibitemShut
  {NoStop}%
\bibitem [{\citenamefont {Perram}\ and\ \citenamefont
  {Wertheim}(1985)}]{perram1985contact}%
  \BibitemOpen
  \bibfield  {author} {\bibinfo {author} {\bibfnamefont {J.~W.}\ \bibnamefont
  {Perram}}\ and\ \bibinfo {author} {\bibfnamefont {M.~S.}\ \bibnamefont
  {Wertheim}},\ }\href {https://doi.org/10.1016/0021-9991(85)90171-8}
  {\bibfield  {journal} {\bibinfo  {journal} {Journal of Computational
  Physics}\ }\textbf {\bibinfo {volume} {58}},\ \bibinfo {pages} {409}
  (\bibinfo {year} {1985})}\BibitemShut {NoStop}%
\bibitem [{\citenamefont {Donev}\ \emph {et~al.}(2007)\citenamefont {Donev},
  \citenamefont {Connelly}, \citenamefont {Stillinger},\ and\ \citenamefont
  {Torquato}}]{donev2007nonspherical}%
  \BibitemOpen
  \bibfield  {author} {\bibinfo {author} {\bibfnamefont {A.}~\bibnamefont
  {Donev}}, \bibinfo {author} {\bibfnamefont {R.}~\bibnamefont {Connelly}},
  \bibinfo {author} {\bibfnamefont {F.~H.}\ \bibnamefont {Stillinger}},\ and\
  \bibinfo {author} {\bibfnamefont {S.}~\bibnamefont {Torquato}},\ }\href
  {https://doi.org/10.1103/PhysRevE.75.051304} {\bibfield  {journal} {\bibinfo
  {journal} {Phys. Rev. E}\ }\textbf {\bibinfo {volume} {75}},\ \bibinfo
  {pages} {051304} (\bibinfo {year} {2007})}\BibitemShut {NoStop}%
\bibitem [{SM()}]{SM}%
  \BibitemOpen
  \href@noop {} {}\bibinfo {note} {See Supplemental Material at [URL will be
  inserted by publisher] for model details and additional numerical simulation
  results.}\BibitemShut {Stop}%
\bibitem [{\citenamefont {Menon}\ and\ \citenamefont
  {Ramaswamy}(2009)}]{MenonRamaswamy2009}%
  \BibitemOpen
  \bibfield  {author} {\bibinfo {author} {\bibfnamefont {G.~I.}\ \bibnamefont
  {Menon}}\ and\ \bibinfo {author} {\bibfnamefont {S.}~\bibnamefont
  {Ramaswamy}},\ }\href {https://doi.org/10.1103/PhysRevE.79.061108} {\bibfield
   {journal} {\bibinfo  {journal} {Phys. Rev. E}\ }\textbf {\bibinfo {volume}
  {79}},\ \bibinfo {pages} {061108} (\bibinfo {year} {2009})}\BibitemShut
  {NoStop}%
\bibitem [{Note1()}]{Note1}%
  \BibitemOpen
  \bibinfo {note} {The susceptibility is defined by $$ \chi = \protect \frac
  {N}{\epsilon } \left ( \left \langle n_{\protect \mathrm {CB}}^2 \right
  \rangle - \left \langle n_{\protect \mathrm {CB}} \right \rangle ^2 \right )
  $$ where $n_{\protect \mathrm {CB}} = \protect \frac {1}{N}\DOTSB \sum@
  \slimits@ n_{\protect \mathrm {CB}, i}$}\BibitemShut {NoStop}%
\bibitem [{\citenamefont {Kurchan}(1998)}]{Kurchan1998}%
  \BibitemOpen
  \bibfield  {author} {\bibinfo {author} {\bibfnamefont {J.}~\bibnamefont
  {Kurchan}},\ }\href {https://doi.org/10.1088/0305-4470/31/16/003} {\bibfield
  {journal} {\bibinfo  {journal} {Journal of Physics A: Mathematical and
  General}\ }\textbf {\bibinfo {volume} {31}},\ \bibinfo {pages} {3719}
  (\bibinfo {year} {1998})}\BibitemShut {NoStop}%
\bibitem [{\citenamefont {Lebowitz}\ and\ \citenamefont
  {Spohn}(1999)}]{LebowitzSpohn1999}%
  \BibitemOpen
  \bibfield  {author} {\bibinfo {author} {\bibfnamefont {J.~L.}\ \bibnamefont
  {Lebowitz}}\ and\ \bibinfo {author} {\bibfnamefont {H.}~\bibnamefont
  {Spohn}},\ }\href {https://doi.org/10.1023/A:1004589714161} {\bibfield
  {journal} {\bibinfo  {journal} {Journal of Statistical Physics}\ }\textbf
  {\bibinfo {volume} {95}},\ \bibinfo {pages} {333} (\bibinfo {year}
  {1999})}\BibitemShut {NoStop}%
\bibitem [{\citenamefont {Ro}\ \emph {et~al.}(2022)\citenamefont {Ro},
  \citenamefont {Guo}, \citenamefont {Shih}, \citenamefont {Phan},
  \citenamefont {Austin}, \citenamefont {Levine}, \citenamefont {Chaikin},\
  and\ \citenamefont {Martiniani}}]{ro2022model}%
  \BibitemOpen
  \bibfield  {author} {\bibinfo {author} {\bibfnamefont {S.}~\bibnamefont
  {Ro}}, \bibinfo {author} {\bibfnamefont {B.}~\bibnamefont {Guo}}, \bibinfo
  {author} {\bibfnamefont {A.}~\bibnamefont {Shih}}, \bibinfo {author}
  {\bibfnamefont {T.~V.}\ \bibnamefont {Phan}}, \bibinfo {author}
  {\bibfnamefont {R.~H.}\ \bibnamefont {Austin}}, \bibinfo {author}
  {\bibfnamefont {D.}~\bibnamefont {Levine}}, \bibinfo {author} {\bibfnamefont
  {P.~M.}\ \bibnamefont {Chaikin}},\ and\ \bibinfo {author} {\bibfnamefont
  {S.}~\bibnamefont {Martiniani}},\ }\href@noop {} {\bibfield  {journal}
  {\bibinfo  {journal} {Physical Review Letters}\ }\textbf {\bibinfo {volume}
  {129}},\ \bibinfo {pages} {220601} (\bibinfo {year} {2022})}\BibitemShut
  {NoStop}%
\bibitem [{\citenamefont {Wang}\ and\ \citenamefont
  {Lebowitz}(1988)}]{WangLebowitz1988NESSIsing}%
  \BibitemOpen
  \bibfield  {author} {\bibinfo {author} {\bibfnamefont {J.-S.}\ \bibnamefont
  {Wang}}\ and\ \bibinfo {author} {\bibfnamefont {J.~L.}\ \bibnamefont
  {Lebowitz}},\ }\href {https://doi.org/10.1007/BF01014891} {\bibfield
  {journal} {\bibinfo  {journal} {Journal of Statistical Physics}\ }\textbf
  {\bibinfo {volume} {51}},\ \bibinfo {pages} {893} (\bibinfo {year}
  {1988})}\BibitemShut {NoStop}%
\bibitem [{\citenamefont
  {Marques}(1989)}]{Marques1989LocallyCompetingTemperatures}%
  \BibitemOpen
  \bibfield  {author} {\bibinfo {author} {\bibfnamefont {M.~C.}\ \bibnamefont
  {Marques}},\ }\href {https://doi.org/10.1088/0305-4470/22/20/025} {\bibfield
  {journal} {\bibinfo  {journal} {Journal of Physics A: Mathematical and
  General}\ }\textbf {\bibinfo {volume} {22}},\ \bibinfo {pages} {4493}
  (\bibinfo {year} {1989})}\BibitemShut {NoStop}%
\bibitem [{\citenamefont {Blote}\ \emph {et~al.}(1990)\citenamefont {Blote},
  \citenamefont {Heringa}, \citenamefont {Hoogland},\ and\ \citenamefont
  {Zia}}]{Blote1990TwoTemperatures}%
  \BibitemOpen
  \bibfield  {author} {\bibinfo {author} {\bibfnamefont {H.~W.~J.}\
  \bibnamefont {Blote}}, \bibinfo {author} {\bibfnamefont {J.~R.}\ \bibnamefont
  {Heringa}}, \bibinfo {author} {\bibfnamefont {A.}~\bibnamefont {Hoogland}},\
  and\ \bibinfo {author} {\bibfnamefont {R.~K.~P.}\ \bibnamefont {Zia}},\
  }\href {https://doi.org/10.1088/0305-4470/23/16/025} {\bibfield  {journal}
  {\bibinfo  {journal} {Journal of Physics A: Mathematical and General}\
  }\textbf {\bibinfo {volume} {23}},\ \bibinfo {pages} {3799} (\bibinfo {year}
  {1990})}\BibitemShut {NoStop}%
\bibitem [{\citenamefont {Tamayo}\ \emph {et~al.}(1994)\citenamefont {Tamayo},
  \citenamefont {Alexander},\ and\ \citenamefont
  {Gupta}}]{Tamayo1994TwoTemperatureIsing}%
  \BibitemOpen
  \bibfield  {author} {\bibinfo {author} {\bibfnamefont {P.}~\bibnamefont
  {Tamayo}}, \bibinfo {author} {\bibfnamefont {F.~J.}\ \bibnamefont
  {Alexander}},\ and\ \bibinfo {author} {\bibfnamefont {R.}~\bibnamefont
  {Gupta}},\ }\href {https://doi.org/10.1103/PhysRevE.50.3474} {\bibfield
  {journal} {\bibinfo  {journal} {Physical Review E}\ }\textbf {\bibinfo
  {volume} {50}},\ \bibinfo {pages} {3474} (\bibinfo {year}
  {1994})}\BibitemShut {NoStop}%
\bibitem [{\citenamefont {Achahbar}\ \emph {et~al.}(1996)\citenamefont
  {Achahbar}, \citenamefont {Alonso},\ and\ \citenamefont
  {Mu{\~n}oz}}]{Achahbar1996NonequilibriumIsing}%
  \BibitemOpen
  \bibfield  {author} {\bibinfo {author} {\bibfnamefont {A.}~\bibnamefont
  {Achahbar}}, \bibinfo {author} {\bibfnamefont {J.~J.}\ \bibnamefont
  {Alonso}},\ and\ \bibinfo {author} {\bibfnamefont {M.~A.}\ \bibnamefont
  {Mu{\~n}oz}},\ }\href {https://doi.org/10.1103/PhysRevE.54.4838} {\bibfield
  {journal} {\bibinfo  {journal} {Physical Review E}\ }\textbf {\bibinfo
  {volume} {54}},\ \bibinfo {pages} {4838} (\bibinfo {year}
  {1996})}\BibitemShut {NoStop}%
\bibitem [{\citenamefont {de~Oliveira}(1992)}]{Oliveira1992MajorityVote}%
  \BibitemOpen
  \bibfield  {author} {\bibinfo {author} {\bibfnamefont {M.~J.}\ \bibnamefont
  {de~Oliveira}},\ }\href {https://doi.org/10.1007/BF01060069} {\bibfield
  {journal} {\bibinfo  {journal} {Journal of Statistical Physics}\ }\textbf
  {\bibinfo {volume} {66}},\ \bibinfo {pages} {273} (\bibinfo {year}
  {1992})}\BibitemShut {NoStop}%
\bibitem [{\citenamefont {de~Oliveira}\ \emph {et~al.}(1993)\citenamefont
  {de~Oliveira}, \citenamefont {Mendes},\ and\ \citenamefont
  {Santos}}]{OliveiraMendesSantos1993IsingUniversal}%
  \BibitemOpen
  \bibfield  {author} {\bibinfo {author} {\bibfnamefont {M.~J.}\ \bibnamefont
  {de~Oliveira}}, \bibinfo {author} {\bibfnamefont {J.~F.~F.}\ \bibnamefont
  {Mendes}},\ and\ \bibinfo {author} {\bibfnamefont {M.~A.}\ \bibnamefont
  {Santos}},\ }\href {https://doi.org/10.1088/0305-4470/26/10/006} {\bibfield
  {journal} {\bibinfo  {journal} {Journal of Physics A: Mathematical and
  General}\ }\textbf {\bibinfo {volume} {26}},\ \bibinfo {pages} {2317}
  (\bibinfo {year} {1993})}\BibitemShut {NoStop}%
\bibitem [{\citenamefont {Marro}\ \emph {et~al.}(1994)\citenamefont {Marro},
  \citenamefont {Fern{\'a}ndez}, \citenamefont {Gonz{\'a}lez-Miranda},\ and\
  \citenamefont {Puma}}]{Marro1994NonHamiltonian}%
  \BibitemOpen
  \bibfield  {author} {\bibinfo {author} {\bibfnamefont {J.}~\bibnamefont
  {Marro}}, \bibinfo {author} {\bibfnamefont {J.~F.}\ \bibnamefont
  {Fern{\'a}ndez}}, \bibinfo {author} {\bibfnamefont {J.~M.}\ \bibnamefont
  {Gonz{\'a}lez-Miranda}},\ and\ \bibinfo {author} {\bibfnamefont
  {M.}~\bibnamefont {Puma}},\ }\href {https://doi.org/10.1103/PhysRevE.50.3237}
  {\bibfield  {journal} {\bibinfo  {journal} {Physical Review E}\ }\textbf
  {\bibinfo {volume} {50}},\ \bibinfo {pages} {3237} (\bibinfo {year}
  {1994})}\BibitemShut {NoStop}%
\bibitem [{\citenamefont {Mendes}\ and\ \citenamefont
  {Santos}(1998)}]{MendesSantos1998ShortTimeMajorityVote}%
  \BibitemOpen
  \bibfield  {author} {\bibinfo {author} {\bibfnamefont {J.~F.~F.}\
  \bibnamefont {Mendes}}\ and\ \bibinfo {author} {\bibfnamefont {M.~A.}\
  \bibnamefont {Santos}},\ }\href {https://doi.org/10.1103/PhysRevE.57.108}
  {\bibfield  {journal} {\bibinfo  {journal} {Physical Review E}\ }\textbf
  {\bibinfo {volume} {57}},\ \bibinfo {pages} {108} (\bibinfo {year}
  {1998})}\BibitemShut {NoStop}%
\bibitem [{\citenamefont {Vieira}\ and\ \citenamefont
  {Crokidakis}(2016)}]{VieiraCrokidakis2016TwoNoises}%
  \BibitemOpen
  \bibfield  {author} {\bibinfo {author} {\bibfnamefont {A.~R.}\ \bibnamefont
  {Vieira}}\ and\ \bibinfo {author} {\bibfnamefont {N.}~\bibnamefont
  {Crokidakis}},\ }\href {https://doi.org/10.1016/j.physa.2016.01.013}
  {\bibfield  {journal} {\bibinfo  {journal} {Physica A: Statistical Mechanics
  and its Applications}\ }\textbf {\bibinfo {volume} {450}},\ \bibinfo {pages}
  {30} (\bibinfo {year} {2016})}\BibitemShut {NoStop}%
\bibitem [{\citenamefont {Vilela}\ and\ \citenamefont
  {Stanley}(2018)}]{VilelaStanley2018StrongOpinions}%
  \BibitemOpen
  \bibfield  {author} {\bibinfo {author} {\bibfnamefont {A.~L.~M.}\
  \bibnamefont {Vilela}}\ and\ \bibinfo {author} {\bibfnamefont {H.~E.}\
  \bibnamefont {Stanley}},\ }\href {https://doi.org/10.1038/s41598-018-26919-y}
  {\bibfield  {journal} {\bibinfo  {journal} {Scientific Reports}\ }\textbf
  {\bibinfo {volume} {8}},\ \bibinfo {pages} {8709} (\bibinfo {year}
  {2018})}\BibitemShut {NoStop}%
\bibitem [{\citenamefont {Korniss}\ \emph {et~al.}(2000)\citenamefont
  {Korniss}, \citenamefont {White}, \citenamefont {Rikvold},\ and\
  \citenamefont {Novotny}}]{Korniss2000DynamicPhaseTransition}%
  \BibitemOpen
  \bibfield  {author} {\bibinfo {author} {\bibfnamefont {G.}~\bibnamefont
  {Korniss}}, \bibinfo {author} {\bibfnamefont {C.~J.}\ \bibnamefont {White}},
  \bibinfo {author} {\bibfnamefont {P.~A.}\ \bibnamefont {Rikvold}},\ and\
  \bibinfo {author} {\bibfnamefont {M.~A.}\ \bibnamefont {Novotny}},\ }\href
  {https://doi.org/10.1103/PhysRevE.63.016120} {\bibfield  {journal} {\bibinfo
  {journal} {Physical Review E}\ }\textbf {\bibinfo {volume} {63}},\ \bibinfo
  {pages} {016120} (\bibinfo {year} {2000})}\BibitemShut {NoStop}%
\bibitem [{\citenamefont {Buend{\'i}a}\ and\ \citenamefont
  {Rikvold}(2008)}]{BuendiaRikvold2008SoftDynamics}%
  \BibitemOpen
  \bibfield  {author} {\bibinfo {author} {\bibfnamefont {G.~M.}\ \bibnamefont
  {Buend{\'i}a}}\ and\ \bibinfo {author} {\bibfnamefont {P.~A.}\ \bibnamefont
  {Rikvold}},\ }\href {https://doi.org/10.1103/PhysRevE.78.051108} {\bibfield
  {journal} {\bibinfo  {journal} {Physical Review E}\ }\textbf {\bibinfo
  {volume} {78}},\ \bibinfo {pages} {051108} (\bibinfo {year}
  {2008})}\BibitemShut {NoStop}%
\bibitem [{\citenamefont {Vatansever}\ and\ \citenamefont
  {Fytas}(2018)}]{VatanseverFytas2018QuenchedRandomness}%
  \BibitemOpen
  \bibfield  {author} {\bibinfo {author} {\bibfnamefont {E.}~\bibnamefont
  {Vatansever}}\ and\ \bibinfo {author} {\bibfnamefont {N.~G.}\ \bibnamefont
  {Fytas}},\ }\href {https://doi.org/10.1103/PhysRevE.97.062146} {\bibfield
  {journal} {\bibinfo  {journal} {Physical Review E}\ }\textbf {\bibinfo
  {volume} {97}},\ \bibinfo {pages} {062146} (\bibinfo {year}
  {2018})}\BibitemShut {NoStop}%
\bibitem [{\citenamefont {Vatansever}\ \emph {et~al.}(2024)\citenamefont
  {Vatansever}, \citenamefont {Vatansever}, \citenamefont {Berger},
  \citenamefont {Vasilopoulos},\ and\ \citenamefont
  {Fytas}}]{Vatansever2024NonantisymmetricField}%
  \BibitemOpen
  \bibfield  {author} {\bibinfo {author} {\bibfnamefont {Z.~D.}\ \bibnamefont
  {Vatansever}}, \bibinfo {author} {\bibfnamefont {E.}~\bibnamefont
  {Vatansever}}, \bibinfo {author} {\bibfnamefont {A.}~\bibnamefont {Berger}},
  \bibinfo {author} {\bibfnamefont {A.}~\bibnamefont {Vasilopoulos}},\ and\
  \bibinfo {author} {\bibfnamefont {N.~G.}\ \bibnamefont {Fytas}},\ }\href
  {https://doi.org/10.1103/PhysRevE.110.064155} {\bibfield  {journal} {\bibinfo
   {journal} {Physical Review E}\ }\textbf {\bibinfo {volume} {110}},\ \bibinfo
  {pages} {064155} (\bibinfo {year} {2024})}\BibitemShut {NoStop}%
\bibitem [{\citenamefont {Halder}\ and\ \citenamefont
  {Acharyya}(2016)}]{HalderAcharyya2016StandingWave}%
  \BibitemOpen
  \bibfield  {author} {\bibinfo {author} {\bibfnamefont {A.}~\bibnamefont
  {Halder}}\ and\ \bibinfo {author} {\bibfnamefont {M.}~\bibnamefont
  {Acharyya}},\ }\href {https://doi.org/10.1016/j.jmmm.2016.07.062} {\bibfield
  {journal} {\bibinfo  {journal} {Journal of Magnetism and Magnetic Materials}\
  }\textbf {\bibinfo {volume} {420}},\ \bibinfo {pages} {290} (\bibinfo {year}
  {2016})}\BibitemShut {NoStop}%
\bibitem [{\citenamefont {Halder}\ and\ \citenamefont
  {Acharyya}(2017)}]{HalderAcharyya2017SpinSWaves}%
  \BibitemOpen
  \bibfield  {author} {\bibinfo {author} {\bibfnamefont {A.}~\bibnamefont
  {Halder}}\ and\ \bibinfo {author} {\bibfnamefont {M.}~\bibnamefont
  {Acharyya}},\ }\href {https://doi.org/10.1088/0253-6102/68/5/600} {\bibfield
  {journal} {\bibinfo  {journal} {Communications in Theoretical Physics}\
  }\textbf {\bibinfo {volume} {68}},\ \bibinfo {pages} {600} (\bibinfo {year}
  {2017})}\BibitemShut {NoStop}%
\bibitem [{\citenamefont {Solon}\ and\ \citenamefont
  {Tailleur}(2015)}]{SolonTailleur2015ActiveIsing}%
  \BibitemOpen
  \bibfield  {author} {\bibinfo {author} {\bibfnamefont {A.~P.}\ \bibnamefont
  {Solon}}\ and\ \bibinfo {author} {\bibfnamefont {J.}~\bibnamefont
  {Tailleur}},\ }\href {https://doi.org/10.1103/PhysRevE.92.042119} {\bibfield
  {journal} {\bibinfo  {journal} {Physical Review E}\ }\textbf {\bibinfo
  {volume} {92}},\ \bibinfo {pages} {042119} (\bibinfo {year}
  {2015})}\BibitemShut {NoStop}%
\bibitem [{\citenamefont {Jos\'e}\ \emph {et~al.}(1977)\citenamefont {Jos\'e},
  \citenamefont {Kadanoff}, \citenamefont {Kirkpatrick},\ and\ \citenamefont
  {Nelson}}]{PhysRevB.16.1217}%
  \BibitemOpen
  \bibfield  {author} {\bibinfo {author} {\bibfnamefont {J.~V.}\ \bibnamefont
  {Jos\'e}}, \bibinfo {author} {\bibfnamefont {L.~P.}\ \bibnamefont
  {Kadanoff}}, \bibinfo {author} {\bibfnamefont {S.}~\bibnamefont
  {Kirkpatrick}},\ and\ \bibinfo {author} {\bibfnamefont {D.~R.}\ \bibnamefont
  {Nelson}},\ }\href {https://doi.org/10.1103/PhysRevB.16.1217} {\bibfield
  {journal} {\bibinfo  {journal} {Phys. Rev. B}\ }\textbf {\bibinfo {volume}
  {16}},\ \bibinfo {pages} {1217} (\bibinfo {year} {1977})}\BibitemShut
  {NoStop}%
\bibitem [{\citenamefont {Allen}\ and\ \citenamefont
  {Cahn}(1979)}]{ALLEN19791085}%
  \BibitemOpen
  \bibfield  {author} {\bibinfo {author} {\bibfnamefont {S.~M.}\ \bibnamefont
  {Allen}}\ and\ \bibinfo {author} {\bibfnamefont {J.~W.}\ \bibnamefont
  {Cahn}},\ }\href {https://doi.org/10.1016/0001-6160(79)90196-2} {\bibfield
  {journal} {\bibinfo  {journal} {Acta Metallurgica}\ }\textbf {\bibinfo
  {volume} {27}},\ \bibinfo {pages} {1085} (\bibinfo {year}
  {1979})}\BibitemShut {NoStop}%
\end{thebibliography}%


\begin{thebibliography}{5}%
\makeatletter
\providecommand \@ifxundefined [1]{%
 \@ifx{#1\undefined}
}%
\providecommand \@ifnum [1]{%
 \ifnum #1\expandafter \@firstoftwo
 \else \expandafter \@secondoftwo
 \fi
}%
\providecommand \@ifx [1]{%
 \ifx #1\expandafter \@firstoftwo
 \else \expandafter \@secondoftwo
 \fi
}%
\providecommand \natexlab [1]{#1}%
\providecommand \enquote  [1]{``#1''}%
\providecommand \bibnamefont  [1]{#1}%
\providecommand \bibfnamefont [1]{#1}%
\providecommand \citenamefont [1]{#1}%
\providecommand \href@noop [0]{\@secondoftwo}%
\providecommand \href [0]{\begingroup \@sanitize@url \@href}%
\providecommand \@href[1]{\@@startlink{#1}\@@href}%
\providecommand \@@href[1]{\endgroup#1\@@endlink}%
\providecommand \@sanitize@url [0]{\catcode `\\12\catcode `\$12\catcode
  `\&12\catcode `\#12\catcode `\^12\catcode `\_12\catcode `\%12\relax}%
\providecommand \@@startlink[1]{}%
\providecommand \@@endlink[0]{}%
\providecommand \url  [0]{\begingroup\@sanitize@url \@url }%
\providecommand \@url [1]{\endgroup\@href {#1}{\urlprefix }}%
\providecommand \urlprefix  [0]{URL }%
\providecommand \Eprint [0]{\href }%
\providecommand \doibase [0]{https://doi.org/}%
\providecommand \selectlanguage [0]{\@gobble}%
\providecommand \bibinfo  [0]{\@secondoftwo}%
\providecommand \bibfield  [0]{\@secondoftwo}%
\providecommand \translation [1]{[#1]}%
\providecommand \BibitemOpen [0]{}%
\providecommand \bibitemStop [0]{}%
\providecommand \bibitemNoStop [0]{.\EOS\space}%
\providecommand \EOS [0]{\spacefactor3000\relax}%
\providecommand \BibitemShut  [1]{\csname bibitem#1\endcsname}%
\let\auto@bib@innerbib\@empty
\bibitem [{\citenamefont {Perram}\ and\ \citenamefont
  {Wertheim}(1985)}]{perram1985contact}%
  \BibitemOpen
  \bibfield  {author} {\bibinfo {author} {\bibfnamefont {J.~W.}\ \bibnamefont
  {Perram}}\ and\ \bibinfo {author} {\bibfnamefont {M.~S.}\ \bibnamefont
  {Wertheim}},\ }\href {https://doi.org/10.1016/0021-9991(85)90171-8}
  {\bibfield  {journal} {\bibinfo  {journal} {Journal of Computational
  Physics}\ }\textbf {\bibinfo {volume} {58}},\ \bibinfo {pages} {409}
  (\bibinfo {year} {1985})}\BibitemShut {NoStop}%
\bibitem [{\citenamefont {Donev}\ \emph {et~al.}(2007)\citenamefont {Donev},
  \citenamefont {Connelly}, \citenamefont {Stillinger},\ and\ \citenamefont
  {Torquato}}]{donev2007nonspherical}%
  \BibitemOpen
  \bibfield  {author} {\bibinfo {author} {\bibfnamefont {A.}~\bibnamefont
  {Donev}}, \bibinfo {author} {\bibfnamefont {R.}~\bibnamefont {Connelly}},
  \bibinfo {author} {\bibfnamefont {F.~H.}\ \bibnamefont {Stillinger}},\ and\
  \bibinfo {author} {\bibfnamefont {S.}~\bibnamefont {Torquato}},\ }\href
  {https://doi.org/10.1103/PhysRevE.75.051304} {\bibfield  {journal} {\bibinfo
  {journal} {Phys. Rev. E}\ }\textbf {\bibinfo {volume} {75}},\ \bibinfo
  {pages} {051304} (\bibinfo {year} {2007})}\BibitemShut {NoStop}%
\bibitem [{\citenamefont {Kurchan}(1998)}]{Kurchan1998}%
  \BibitemOpen
  \bibfield  {author} {\bibinfo {author} {\bibfnamefont {J.}~\bibnamefont
  {Kurchan}},\ }\href {https://doi.org/10.1088/0305-4470/31/16/003} {\bibfield
  {journal} {\bibinfo  {journal} {Journal of Physics A: Mathematical and
  General}\ }\textbf {\bibinfo {volume} {31}},\ \bibinfo {pages} {3719}
  (\bibinfo {year} {1998})}\BibitemShut {NoStop}%
\bibitem [{\citenamefont {Lebowitz}\ and\ \citenamefont
  {Spohn}(1999)}]{LebowitzSpohn1999}%
  \BibitemOpen
  \bibfield  {author} {\bibinfo {author} {\bibfnamefont {J.~L.}\ \bibnamefont
  {Lebowitz}}\ and\ \bibinfo {author} {\bibfnamefont {H.}~\bibnamefont
  {Spohn}},\ }\href {https://doi.org/10.1023/A:1004589714161} {\bibfield
  {journal} {\bibinfo  {journal} {Journal of Statistical Physics}\ }\textbf
  {\bibinfo {volume} {95}},\ \bibinfo {pages} {333} (\bibinfo {year}
  {1999})}\BibitemShut {NoStop}%
\bibitem [{\citenamefont {Ro}\ \emph {et~al.}(2022)\citenamefont {Ro},
  \citenamefont {Guo}, \citenamefont {Shih}, \citenamefont {Phan},
  \citenamefont {Austin}, \citenamefont {Levine}, \citenamefont {Chaikin},\
  and\ \citenamefont {Martiniani}}]{ro2022model}%
  \BibitemOpen
  \bibfield  {author} {\bibinfo {author} {\bibfnamefont {S.}~\bibnamefont
  {Ro}}, \bibinfo {author} {\bibfnamefont {B.}~\bibnamefont {Guo}}, \bibinfo
  {author} {\bibfnamefont {A.}~\bibnamefont {Shih}}, \bibinfo {author}
  {\bibfnamefont {T.~V.}\ \bibnamefont {Phan}}, \bibinfo {author}
  {\bibfnamefont {R.~H.}\ \bibnamefont {Austin}}, \bibinfo {author}
  {\bibfnamefont {D.}~\bibnamefont {Levine}}, \bibinfo {author} {\bibfnamefont
  {P.~M.}\ \bibnamefont {Chaikin}},\ and\ \bibinfo {author} {\bibfnamefont
  {S.}~\bibnamefont {Martiniani}},\ }\href@noop {} {\bibfield  {journal}
  {\bibinfo  {journal} {Physical Review Letters}\ }\textbf {\bibinfo {volume}
  {129}},\ \bibinfo {pages} {220601} (\bibinfo {year} {2022})}\BibitemShut
  {NoStop}%
\end{thebibliography}%
\end{document}